\pdfoutput=1
\RequirePackage{fixltx2e}
\documentclass[aip,jcp,reprint,floatfix]{revtex4-1}

\usepackage{amsmath,amsfonts} %
\usepackage{wasysym} %
\usepackage[bbgreekl]{mathbbol}
\usepackage{graphicx} %
\usepackage[dvipsnames,table]{xcolor} %
\usepackage{tabularx} %
\usepackage[acronym]{glossaries} %
\usepackage{bm} %
\usepackage{braket} %
\usepackage[binary-units=true,detect-weight=true]{siunitx} %
\usepackage{microtype} %
\usepackage{natbib} 
\usepackage{tikz} %
\usepackage[byname]{smartref} %
\usepackage[breaklinks, %
colorlinks, linkcolor=Blue, citecolor=Blue, urlcolor=Blue]{hyperref} %
\usepackage[version=3]{mhchem}
\AtBeginDocument{\usepackage{booktabs}} 

\DeclareMathOperator\diag{diag} %
\newcommand*\conj[1]{{#1}^{*}}

\newcommand{\E}{\textrm{e}}
\newcommand{\I}{\mathrm{i}\mkern1mu}
\newcommand*\numcircledtikz[1]{\tikz[baseline=(char.base)]{
    \node[shape=circle,draw,inner sep=0.25pt] (char) {#1};}}
\newcommand{\RR}{\mathbf{R}} %

\def\MyTitle{Mixed quantum-classical dynamics using collective
  electronic variables: A better alternative to electronic friction
  theories} %
\def\MyAuthora{Ilya G. Ryabinkin} %
\def\MyAuthorb{Artur F. Izmaylov} %
\def\MySubject{Nonadiabatic phenomena} %

\hypersetup{ %
  pdftitle={\MyTitle{}}, %
  pdfauthor={\MyAuthora{} and \MyAuthorb{}}, %
  pdfsubject={\MySubject{}}, %
  pdfkeywords={} %
}

\graphicspath{{pics/}}

\newacronym{MQC}{MQC}{mixed quantum-classical} %
\newacronym[longplural={degrees of freedom}, %
firstplural={degrees of freedom (DOF)}, %
plural={DOF} %
]{DOF}{DOF}{degree of freedom} %
\newacronym{DFT}{DFT}{density functional theory} %
\newacronym{NAC}{NAC}{nonadiabatic coupling} %
\newacronym{FSSH}{FSSH}{fewest-switches surface hopping} %
\newacronym{PES}{PES}{potential energy surface} %
\newacronym[longplural={equations of motion}, %
firstplural={equations of motion (EOM)}, %
plural={EOM}]{EOM}{EOM}{equation of motion} %
\newacronym{SI}{SI}{Supplementary Information}

\begin{document}

\title{\MyTitle}

\author{\MyAuthora} %
\author{\MyAuthorb} %
\affiliation{Department of Physical and Environmental Sciences,
  University of Toronto Scarborough, Toronto, Ontario, M1C\,1A4,
  Canada} %
\affiliation{Chemical Physics Theory Group, Department of Chemistry,
  University of Toronto, Toronto, Ontario M5S\,3H6, Canada} %

\date{\today}

\begin{abstract}
  An accurate description of nonadiabatic dynamics of molecular
  species on metallic surfaces poses a serious computational challenge
  associated with a multitude of closely-spaced electronic states. We
  propose a \acrlong{MQC} scheme that addresses this challenge by
  introducing collective electronic variables. These variables are
  defined through analytic block-diagonalization applied to the
  time-dependent Hamiltonian matrix governing the electronic dynamics.
  We compare our scheme with the Ehrenfest approach and with a
  full-memory electronic friction model on a one-dimensional ``adatom
  + atomic chain'' model. Our simulations demonstrate that
  collective-mode dynamics with only few (2--3) electronic variables
  is robust and can describe a variety of situations: from a
  chemisorbed atom on an insulator to an atom on a metallic surface.
  Our molecular model also reveals that the friction approach is prone
  to unpredictable and catastrophic failures.
\end{abstract}

\glsresetall

\pacs{}
\maketitle


First-principles nonadiabatic simulations of large molecular systems,
such as biomolecules, molecular aggregates and crystals are now
routinely possible with the aid of \gls{MQC}
methods,\cite{Tully:1998/fd/407, Doltsinis:2002/jtcc/319,
  Tully:2012/jcp/22A301, Kapral:2015/jpcm/073201, Wang:2016/jpcl/2100,
  Pittner:2009/cp/147, Barbatti:2013/wcms/26, Fischer:2011/jcp/144102}
which treat nuclei classically whereas electrons quantum-mechanically.
Computational efficiency of \gls{MQC} methods hinges upon a large
spacing between electronic states. Only a small subset of strongly
coupled states needs to be considered and their explicit propagation
poses no computational difficulty. The situation becomes more
complicated for nonadiabatic dynamics on metallic and semiconductor
surfaces. Tracing a plethora of electronic states is computationally
intensive.\cite{Saalfrank:2004/coss/334} Even in the simplest
\gls{MQC} Ehrenfest method,\cite{Tully:1998/fd/407,
  Doltsinis:2002/jtcc/319} a representation chosen for the electronic
wave function cannot contain fewer variables than the number of
accessible states, and hence, hundreds or even thousands of electronic
\glspl{DOF} must be considered for molecular dynamics on a metallic
surface. Computing this number of excited states and nonadiabatic
couplings may not be a daunting task \emph{per se}. For example,
within the independent-particle approximation, like the Kohn--Sham
\acrlong{DFT}, the excitation frequencies are only orbital energy
differences,\cite{Savin:1998/CPL/391} and the nonadiabatic couplings
can be computed from derivatives of the corresponding
orbitals.\cite{Craig:2005/prl/163001,
  Shenvi:2012/fd/325,Ryabinkin:2015/jpcl/4200} A challenge, therefore,
is how to avoid solving the time-dependent Schr\"odinger equation for
all electronic variables.

This problem has already received some attention. An effective
electronic picture has been proposed by Shenvi {\it et
  al.}\cite{Shenvi:2009/jcp/174107} for the \gls{FSSH} method. A
metallic continuum was discretized through a Gaussian quadrature,
whose nodes and weights were derived from a molecule-metal coupling
(hybridization) function.\cite{Shenvi:2008/pra/022502} Several dozens
of states were sufficient for converged dynamics at a sub-picosecond
timescale.\cite{Shenvi:2009/jcp/174107} Despite all the merits, this
approach requires prior knowledge of the hybridization function, which
has to be deduced from the Newns--Anderson model
Hamiltonian,\cite{Newns:1969/pr/1123} making this approach unsuitable
for on-the-fly simulations.

More traditional approach to systems with a dense electronic spectrum
is to eliminate the explicit electronic dynamics entirely. Typically,
the electronic subsystem is described as a bath of harmonic
oscillators. However, realistic electron-nuclear couplings do not
allow for analytic integration of the electronic \glspl{EOM}, and
further simplifications become inevitable. These usually involve the
constant-coupling approximation, which leads to the Markovian limit,
and an assumption that the density of states around the Fermi level is
smooth.\cite{Head-Gordon:1995/jcp/10137} Several such electronic
friction theories have been presented over the
years.\cite{Persson:1982/prl/662, Hellsing:1984/ps/360,
  Head-Gordon:1995/jcp/10137, Echenique:1981/ssc/779,
  Puska:1983/prb/6121, Juaristi:1999/prl/1048,
  Askerka:2016/prl/217601, Dou:2015/jcp/054103, Dou:2017/jcp/092304}
All of them were capable to capture experimental trends when applied
to selected realistic systems.\cite{Trail:2002/prl/166802,
  Trail:2003/jcp/4539, Head-Gordon:1992/jcp/3939,
  Head-Gordon:1992/prb/1853, Tully:1993/jvs/1914, Kindt:1998/jcp/3629,
  Juaristi:2008/prl/116102, Shenvi:2009/jcp/174107,
  Monturet:2010/prb/075404, Fuchsel:2011/pccp/8659,
  BlancoRey:2014/prl/103203, Rittmeyer:2015/prl/046102,
  Novko:2015/prb/201411, Askerka:2016/prl/217601}

Surprisingly, none of these frictional theories have been scrutinized
on models where the Ehrenfest method can be used to gauge the adequacy
of introduced dynamical approximations. To the best of our knowledge,
only Ref.~\citenum{Shenvi:2009/jcp/174107} made a direct comparison of
the \citet{Head-Gordon:1995/jcp/10137} friction model with the
\gls{FSSH}-style dynamics. Authors point out that the friction model
predicts somewhat faster relaxation at short times
($t < \SI{150}{\femto\second}$) as compared to its surface-hopping
counterpart. There are other indications that friction models may
exaggerate nuclear energy dissipation: for example,
Refs.~\citenum{Trail:2002/prl/166802}
and~\citenum{Trail:2003/jcp/4539} report the divergence of the
friction coefficient (``infinite stopping power'') for the \ce{H} on
\ce{Cu}(111) system. However, it is not clear to what level of
approximation these issues must be attributed.

In this Letter we carefully assess the friction approximation and
propose a computationally superior scheme that employs a
superadiabatic-like transformation.\cite{Deschamps:2008/jcp/204110,
  Lim:1991/jpa/3255} This transformation introduces approximate
dynamical decoupling of electronic \glspl{DOF} and allows us to reduce
the number of retained electronic variables to only few. With the new
formalism we aim at the description of systems with a dense but
potentially highly non-uniform electronic spectrum, such as chemi- and
physisorbed atoms, radicals, or molecules on metallic
surfaces.\cite{vanSanten:2016/pccp/1}


{\it Ehrenfest and electronic friction:} The Ehrenfest method can be
derived\cite{Minnhagen:1982/jpc/2293, Head-Gordon:1995/jcp/10137} from
a semi-classical energy functional
\begin{equation}
  \label{eq:W}
  W =  \sum_\alpha \frac{M_\alpha {\dot
      R}^2_\alpha}{2} + \braket{\Psi|\hat H_e|\Psi},
\end{equation}
where $\mathbf{R} \equiv \{R_\alpha\}$ is a collection of nuclear
classical \glspl{DOF}, a dot stands for the full time derivative,
$\ket{\Psi}$ is a time-dependent electronic wave function, and
$\hat H_e$ is the electronic Hamiltonian which operates on electronic
variables and is a parametric function of $\mathbf{R}$. In the
interaction picture $\ket{\Psi}$ is\cite{Doltsinis:2002/jtcc/319}:
\begin{equation}
  \label{eq:Psi-param}
  \ket{\Psi} = \sum_j c_j(t) \E^{-\I\phi_j(t)} \ket{\Phi_j[\RR(t)]},
\end{equation}
where
$\phi_j(t) = \hbar^{-1}\int_0^t
E_j\bigl[\mathbf{R}(t')\bigr]\,\mathrm{d}t'$
are dynamical phases, $\{E_j[\mathbf{R}(t)]\}$ and
$\{\ket{\Phi_j[\RR(t)]}\}$ are adiabatic \glspl{PES} and wave
functions, respectively,
$\hat H_e \ket{\Phi_j[\RR(t)]} = E_j[\RR(t)] \ket{\Phi_j[\RR(t)]}$,
and $\{c_j(t)\}$ are complex time-dependent coefficients. Using the
electronic time-dependent Schr\"odinger equation projected onto
$\{\ket{\Phi_j[\RR(t)]}\}$, one can derive the electronic \glspl{EOM}:
\begin{equation}
  \label{eq:ehr_work_el}
  \dot c_k = -\sum_{j} c_j \E^{\I\phi_{kj}}
  \tau_{kj},
\end{equation}
where $\tau_{kj}(t) = \Braket{\Phi_k [\RR(t)]|\dot{\Phi}_j[\RR(t)]}$
are the time-derivative nonadiabatic couplings. Conservation of $W$,
$\dot W = 0$, gives the nuclear
\glspl{EOM}\cite{Doltsinis:2002/jtcc/319}:
\begin{equation}
  \label{eq:ehr_nuc}
  M_\alpha \ddot R_\alpha +\sum_k |c_k(t)|^2 \frac{\partial
    E_k}{\partial R_\alpha} = \sum_{k\ne j} \conj{c}_kc_j
  \E^{\I\phi_{kj}} f^\alpha_{kj},  
\end{equation}
where
$f_{kj}^\alpha(t) = -\Braket{\Phi_k|\partial_{R_\alpha}\hat
  H_e|\Phi_j}$
are nonadiabatic forces along the trajectory and
$\phi_{kj}(t) = \phi_k(t) - \phi_j(t)$.

Two approximations are used for deriving friction theories: 1) the
average force $-\sum_k |c_k(t)|^2 \partial E_k/\partial R_\alpha$ is
replaced with a derivative of the ground-state \gls{PES},
$-\partial E_0/\partial R_\alpha$, in Eq.~\eqref{eq:ehr_nuc}, and 2)
all excited-excited couplings $\tau_{kj}$ are considered to be
insignificant. These approximations are justified by accounting a
predominant role of the ground state and low population of the excited
states. They result in the simplified Ehrenfest equations
\begin{eqnarray}
  \label{eq:ehr_s_el}
  \dot{\mathbf{c}} = -\mathbf{Tc}, \quad T_{kj} = \delta_{0k}\delta_{0j}
  \tau_{kj}\E^{\I\phi_{kj}}, \\
  \label{eq:ehr_s_nuc}
  M_\alpha \ddot R_\alpha + \frac{\partial E_0}{\partial R_\alpha} = 
  \sum_{k\ne j} \conj{c}_kc_j \E^{\I\phi_{kj}} f^\alpha_{kj}.  
\end{eqnarray}

To obtain a microscopic frictional theory we
follow~\citet{Head-Gordon:1995/jcp/10137}. Note, however, that, in
accordance with the original work, two additional approximations have
been adopted: 1) the adiabatic excitation frequencies are
position-independent:
$\hbar\omega_{k0}(\RR)=E_k(\RR)-E_0(\RR)=\text{const}_k$, 2) the
ground electronic state is weakly coupled to the excited ones, which
translates into the condition $|c_0(t)|^2\approx 1$. Under these
assumptions the electronic dynamics can be integrated out, and
Eqs.~\eqref{eq:ehr_s_el} and \eqref{eq:ehr_s_nuc} are converted into
\begin{equation}
  \label{eq:fric}
  M_\alpha \ddot R_\alpha + \frac{\partial E_0}{\partial R_\alpha}
  =  \sum_{\beta} \int_0^t \mathcal{D}^{\alpha\beta}(t,t') \dot
  R_\beta(t')\,\mathrm{d}t' + \mathcal{R}^\alpha(t),
\end{equation}
where
$\mathcal{D}^{\alpha\beta}(t, t') = -2\hbar \sum_{k >0}
f^\alpha_{k0}(t) f^\beta_{k0}(t')\cos[\omega_{k0} (t' -
t)\bigr]/\omega_{k0}$
is the friction kernel, and $\mathcal{R}^\alpha(t)$ represents a
random force associated with generally unknown initial conditions for
the electronic subsystem. In what follows we omit this term as for a
pure initial electronic state it is zero. We integrate the Langevin
equation~\eqref{eq:fric} exactly; details of the integration scheme
can be found in \gls{SI}.\cite{Supp:ref} It is the most accurate
frictional approach possible because of the absolute minimum number of
introduced approximations. Any other models with additional
approximations can only be more accurate due to fortuitous
cancellation of errors.

{\it Two collective mode (2CM) model:} As an alternative to the
frictional model we propose a different treatment of the simplified
Ehrenfest equations \eqref{eq:ehr_s_el} and \eqref{eq:ehr_s_nuc}. The
$\mathbf{T}$ matrix in Eq.~\eqref{eq:ehr_s_el} can be
block-diagonalized analytically. Block-diagonalization of $\mathbf{T}$
gives $\mathbf{Z} = \mathbf{U^{-1} T U}$, where $\mathbf{Z}$ defines a
canonical form of the interaction matrix, and
$\mathbf{U}=\mathbf{U}(\tau_{ij}, \phi_{ij})$\cite{Supp:ref}. The only
non-zero block of $\mathbf{Z}$ is the upper-left corner,
$\mathbf{Z}_{2\times2}= -\I\sigma_y z$, where $\sigma_y$ is the Pauli
matrix and $z = \left(\sum_{k=1} \tau_{k0}^2\right)^{1/2}$. Thus,
$\mathbf{Z}$ acts non-trivially only within a two-dimensional subspace
of transformed electronic variables
$\mathbf{d} = \mathbf{U}^{-1} \mathbf{c}$, where it couples only $d_0$
and $d_1$. We denote this two-dimensional subspace as the
\emph{interaction} space. $d_0$ coincides with the probability
amplitude of the ground adiabatic state $c_0$ and will be used to
track the ground state population evolution. The remaining variables
$\{d_i\}_{i=2, \ldots}$ define the \emph{spectator} space.

The transformation matrix $\mathbf{U}$ is time-dependent, therefore,
the electronic \glspl{EOM} in new variables,
$\dot{\mathbf{d}} = -\left(\mathbf{U^{-1}\dot{U}} +
  \mathbf{Z}\right)\mathbf{d}$,
acquire a new nonadiabatic term $\mathbf{N}=\mathbf{U^{-1}\dot{U}}$.
$\mathbf{N}$ has a quite involved explicit form, but its $2\times2$
upper block is quite simple.\cite{Supp:ref} Therefore, we introduce
our first approximate 2CM method by restricting the electronic
dynamics to the interaction space only
\begin{eqnarray}
  \label{eq:two_mode_el}
  \begin{pmatrix}
    {\dot d}_0 \\
    {\dot d}_1
  \end{pmatrix}
& =
  \begin{pmatrix}
    0 & -z \\
    z & \I{\widetilde w}_{11}
  \end{pmatrix}
        \begin{pmatrix}
          {d}_0 \\
          {d}_1
        \end{pmatrix},
\end{eqnarray}
where ${\widetilde w}_{11} = z^{-2}\sum_{k=1} \omega_{k0}\tau_{k0}^2$
is the collective frequency, which originated from $\mathbf{N}$. The
nonadiabatic force matrix $\{f^\alpha_{kj}(t)\}_{k,j \ge 0}$,
projected onto the interaction space,
$\left\{\mathbf{U}^\dagger \mathbf{f}^\alpha
  \mathbf{U}\right\}_{2\times2}$,
has only one non-zero element,
${\tilde f}_{10}^\alpha = z^{-1} \sum_{k=1} f_{k0}^\alpha\tau_{k0}.$
Using this projection we re-write the nuclear Ehrenfest
\glspl{EOM}~\eqref{eq:ehr_s_nuc} as
\begin{equation}
  \label{eq:two_mode_nuc}
  M_\alpha \ddot R_\alpha + \frac{\partial E_0}{\partial R_\alpha} =
  \left(\conj{d}_0d_1 + \conj{d}_1d_0 \right) {\tilde f}^\alpha_{10}. 
\end{equation}

{\it Three collective mode (3CM) model:} The 2CM model treats all the
original electronic \glspl{DOF} in a democratic fashion in the
collective frequency $\tilde{\omega}_{11}$ and coupling $z$. For a
chemisorbed species on a surface, however, one coupling, which
corresponds to molecular binding, can be appreciably larger than
others. Interaction-space variables in this case may be biased toward
a description of that particular electronic \glspl{DOF}, and the 2CM
dynamics will miss contributions from weakly-coupled states. One
needs, therefore, a procedure to select the most important \glspl{DOF}
from the spectator space. The spectator space interacts with
$(d_0,\, d_1)$-space only through the matrix $\mathbf{N}$. In the
presence of one dominant coupling, without restricting generality we
can take it as $\tau_{10}$, the ratios
$p_{k} = \tau_{k0}/\tau_{10},\ k >1$ and their time derivatives
${\dot p}_{k}$ are small. Therefore, the matrix $\mathbf{N}$ can be
simplified by neglecting $\dot{p}_k$ and off-diagonal quadratic terms
$p_k p_j$:
\begin{align}
  \label{eq:N_3_mode}
  \mathbf{N} & \approx \mathbf{N}^{(a)} = \I\diag{\{0, {\widetilde
               w}_{11}, {\widetilde w}_{22}, \ldots \}} - \I
               \begin{pmatrix}
                 0      & 0               & \mathbf{0} \\
                 0      & 0               & \boldsymbol{\pi} \\
                 \mathbf{0}^\dagger & \boldsymbol{\pi}^{\dagger} &
                 \mathbf{0}
               \end{pmatrix},
\end{align}
where ${\widetilde w}_{kk} = w_{k0} - \tau_{k0}^2\omega_{k1}/z^2$,
$k>1$ and $\boldsymbol{\pi} = (\pi_2, \ldots)$ with
$\pi_k = p_{k}\,\tau_{10}^2 \omega_{k1}/z^2$. In the interaction
representation for $\mathbf{N}^{(a)}$ the diagonal matrix of
collective frequencies [the first term in Eq.~\eqref{eq:N_3_mode}] is
transferred to the corresponding dynamical phases. This reveals the
structure of $\mathbf{N}^{(a)}$, which is similar to that of the
matrix $\mathbf{T}$ [\emph{cf.} Eq.~\eqref{eq:ehr_s_el}] with only
zero first row and column. Thus, we repeat the block-diagonalizing
transformation and separate one additional mode from the spectator
space to obtain a new 3CM model with the following \glspl{EOM}:
\begin{align}
  \label{eq:three_mode_nuc}
  M_\alpha \ddot R_\alpha + \frac{\partial E_0}{\partial R_\alpha} & =
                                \mathbf{d}^\dagger
                                \begin{pmatrix}
                                  0 & {\tilde f}_{10}^\alpha & {\tilde f}_{20}^\alpha \\
                                  {\tilde f}_{10}^\alpha & 0 & 0 \\
                                  {\tilde f}_{20}^\alpha & 0 & 0
                                \end{pmatrix}
  \mathbf{d}, \\
  \label{eq:three_mode_el}
  \dot{\mathbf{d}} & =
                     \begin{pmatrix}
                       0 & -z & 0\\
                       z & \I{\widetilde w}_{11} & -\I\tilde{z} \\
                       0 & -\I\tilde{z} & \I{\widehat w}_{22}
                     \end{pmatrix}
                                          \mathbf{d},
\end{align}
where $\mathbf{d} = (d_0, d_1, d_2)$ are new collective electronic
variables, $\tilde{z} = \left(\sum_{k=2} \pi_k^2\right)^{1/2}$,
${\widehat w}_{22} = {\tilde z}^{-2} \sum_{k=2} {\widetilde w}_{kk}
\pi_k^2$,
and
${\tilde{f}}_{20}^\alpha = {\tilde z}^{-1} z^{-1} \sum_{k=2}
(f_{10}^\alpha\tau_{k0} - f_{k0}^\alpha \tau_{10})\pi_k$.

Recursive nature of the procedure should be apparent now. One can
proceed by separating more modes from the spectator space to obtain
higher-level collective-mode models. The appearance of combination
frequencies [\textit{e.g.}\ $\omega_{k1} = (E_k - E_1)/\hbar$]
warrants, however, that neglecting the excited-excited couplings,
which are dynamical partners of these frequencies, is no longer
justified. Moreover, the transformed forces
${\tilde f}^\alpha_{k0} = 0$, $k \ge 3$, and hence, higher collective
modes do not contribute to the nuclear dynamics directly, although
they are still coupled in the electronic space. That is, the 3CM model
constitutes a natural stopping point for the collective-mode
transformation.

\begin{figure}
  \centering
  \includegraphics[width=0.5\textwidth]{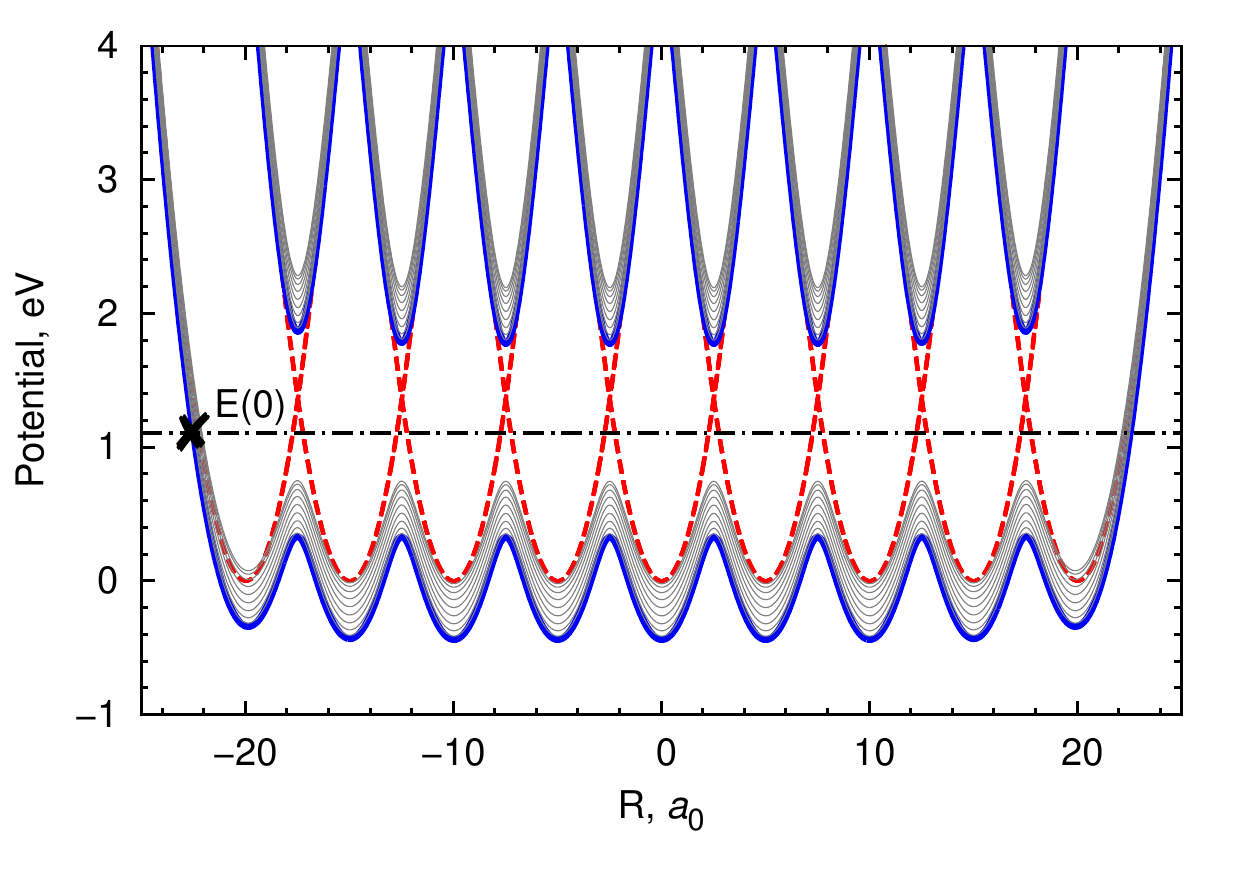}
  \caption{Potential energy surfaces for an adatom-metallic chain
    model, 10 states per layer, other parameters are given in
    \protect\gls{SI}.\cite{Supp:ref} The lowest-energy adiabatic
    \protect\glspl{PES} for each layer (including the ground-state
    one) are solid blue thick lines. Adiabatic \protect\glspl{PES}
    within a layer are solid thin gray lines. Diabatic
    \protect\glspl{PES} are dashed red lines. $E(0)$ is the initial
    energy, a black cross marks the initial position for the
    trajectory.}
  \label{fig:pot_surf}
\end{figure}

{\it Molecular model:} We assess different nonadiabatic approaches on
a model with several electronic and a single nuclear \gls{DOF}. The
model is intended to represent an atom chemisorbed on a
one-dimensional chain of atoms modelling the surface. We parametrize
the electronic Hamiltonian $\hat H_e$ in the \emph{diabatic}
representation as
$\hat H_e = \sum_{ij}\ket{\varphi_i}V_{ij}(R)\bra{\varphi_j}$, where
$V_{ij}(R)$ are either constants for $i\ne j$ or harmonic potentials
(all definitions are detailed in \gls{SI}.\cite{Supp:ref})
All electronic states are organized in nine layers, so that the
off-diagonal couplings are large for states in different layers and
small for states within the same layer. Diagonalization of $\hat H_e$
determines the adiabatic potentials $\{E_k(R)\}$, nonadiabatic
couplings $\{\tau_{kj}(R)\}$ and forces $\{f_{kj}(R)\}$. The resulting
adiabatic \glspl{PES} in a multiple-state-per-layer case are sketched
in Fig.~\ref{fig:pot_surf}. $\tau_{kj}$ and $f_{kj}$ are set to zero
once both $k,j \ge 1$ simultaneously. The nuclear kinetic energy is
$T_N = M{\dot R}^2/2$ with $M = \SI{2000}{\electronmass}$, multiplied
by the identity matrix of the appropriate dimension.
\begin{figure}
  \centering
  \includegraphics[width=1.0\linewidth]{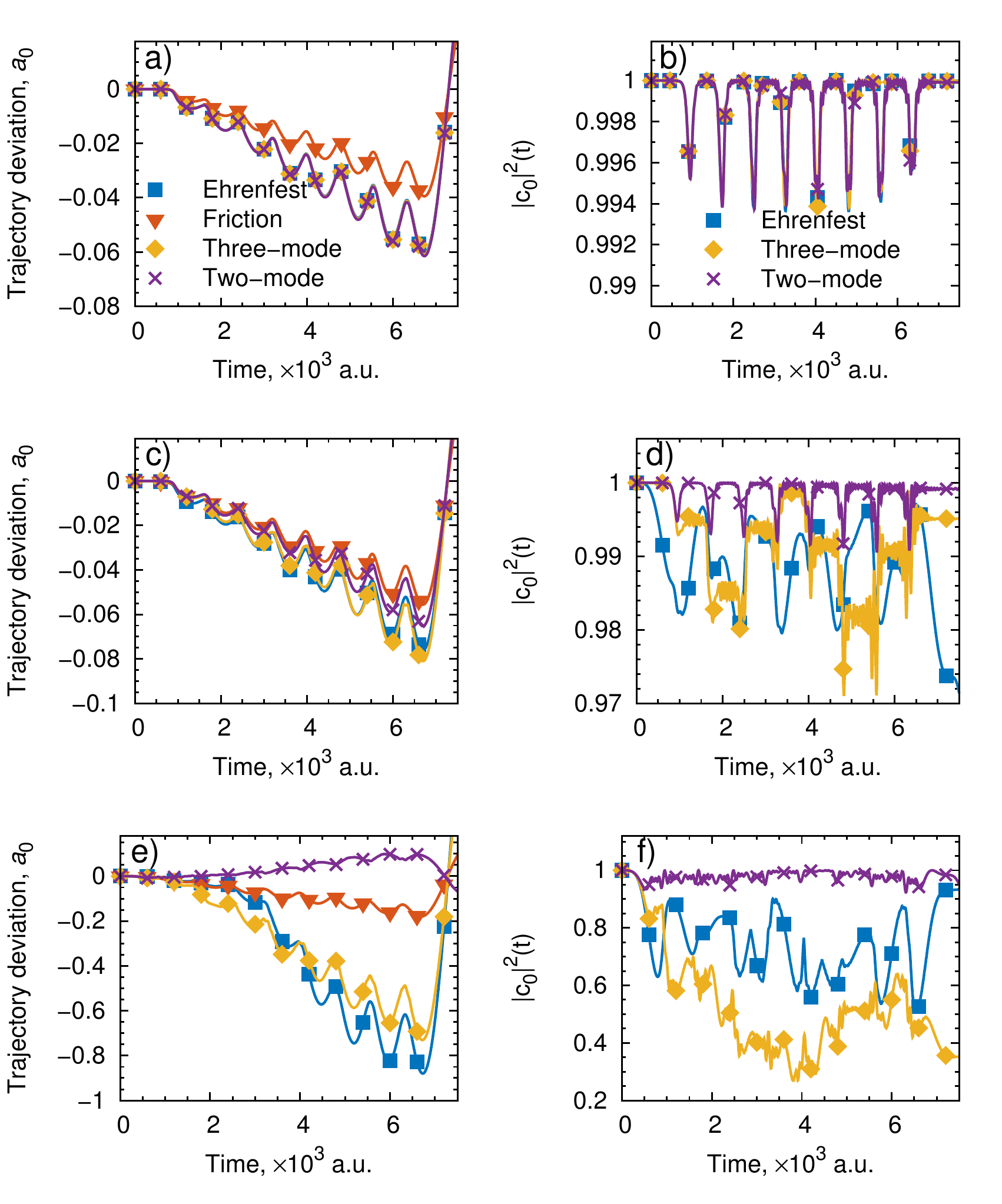}
  \caption{Trajectory dynamics in low-energy, weak-coupling regime.
    Top row (a,b): one state per layer. Middle row (c,d): 5 states per
    layer. Bottom row (e-f): 10 states per layer (see
    Fig.~\ref{fig:pot_surf}). The left column shows deviations of
    trajectories in different methods from that of the
    Born--Oppenheimer approximation, the right column displays
    ground-state weight $|c_0|^2(t)$ dynamics.}
  \label{fig:surf_dyn_1}
\end{figure}

\begin{figure*}
  \centering
  \includegraphics[width=1.0\linewidth]{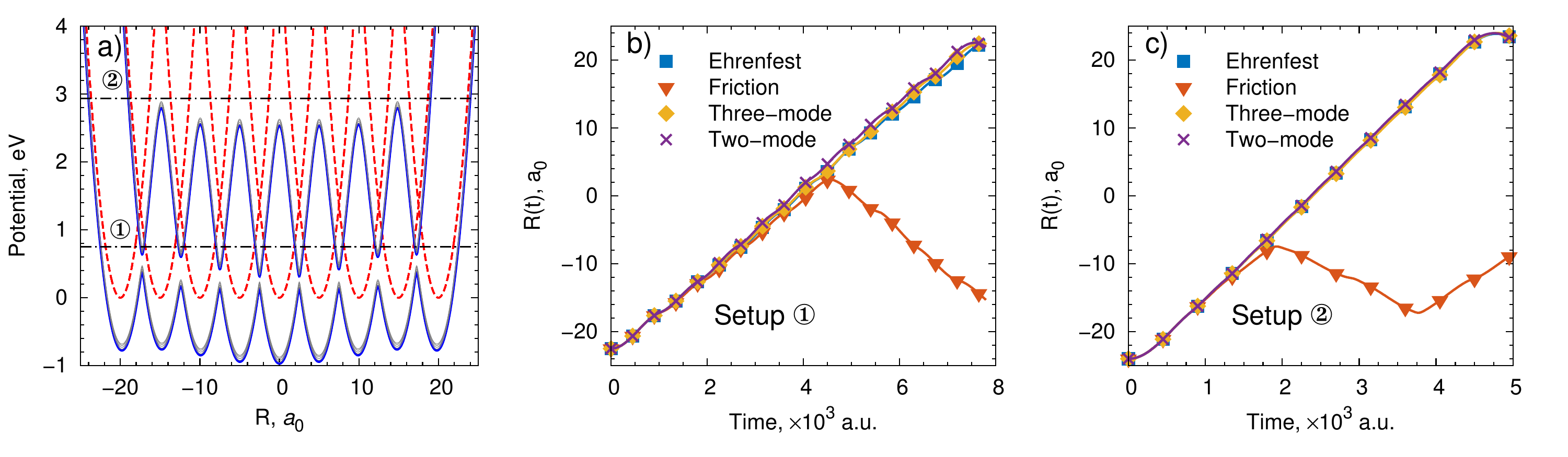}
  \caption{Molecular model has 5 levels per layer, all other
    parameters detailed in \protect\gls{SI}.\cite{Supp:ref} Panel a):
    Adiabatic and diabatic potentials (same as in
    Fig.~\ref{fig:pot_surf}). Horizontal lines labelled as
    \protect\numcircledtikz{\footnotesize 1} and
    \protect\numcircledtikz{\footnotesize 2} show two different
    initial energies. Trajectories for the setups
    \protect\numcircledtikz{\footnotesize 1} and
    \protect\numcircledtikz{\footnotesize 2} are on panels b) and c),
    respectively. }
  \label{fig:fric_fail}
\end{figure*}
{\it Results:} We consider five dynamical approaches: the Ehrenfest
[Eqs.~\eqref{eq:ehr_s_el} and \eqref{eq:ehr_s_nuc}], the full-memory
friction [Eq.~\eqref{eq:fric}], 2CM [Eqs.~\eqref{eq:two_mode_el} and
\eqref{eq:two_mode_nuc}], 3CM [Eqs.~\eqref{eq:three_mode_nuc} and
\eqref{eq:three_mode_el}] models, and the Born-Oppenheimer
approximation [Eq.~\eqref{eq:ehr_s_nuc} with the zero right hand
side]. We start with a set of molecular parameters that correspond to
weak nonadiabatic coupling between layers.\cite{Supp:ref} The initial
energy is taken to be low so that the nonadiabatic dynamics takes
place within a ground-state layer (Fig.~\ref{fig:pot_surf}). The
nonadiabatic effects are small overall: trajectory deviations from
that of the Born-Oppenheimer approximation are fractions of
bohr~(\si{\bohr}) as compared to the total trajectory length of
\SI{50}{\bohr} (Fig.~\ref{fig:surf_dyn_1}).

With one state per layer---an insulator-like surface---both collective
mode models work very well reproducing almost exactly the Ehrenfest
dynamics. Despite the fact that this setup favors the friction
approach as the electronic wave function remains close to the ground
state, the friction model still deviates, since it accounts for
different excitations mostly by the strength of the corresponding
couplings but not according to their actual importance determined by
available energy.

With more states per layer the ground state depletion becomes more
pronounced (Figs.~\ref{fig:surf_dyn_1}d and ~\ref{fig:surf_dyn_1}f).
For 10 states per layer the ground-state weight drops to \emph{ca.}
\SI{50}{\percent}, and the 2CM model breaks down due to the large
number of energetically accessible (albeit weakly-coupled) electronic
states. On the other hand, the 3CM model displays rather accurate
electronic and very accurate nuclear dynamics, as can been seen in
Figs.~\ref{fig:surf_dyn_1}f and \ref{fig:surf_dyn_1}e, respectively.
The nuclear dynamics is accurate because all the \glspl{PES} within a
layer have almost identical profiles, and hence, the nucleus feels the
same force, so that the electronic population within a layer needs to
be accurate only on average. The friction model shows moderate
deviations from the Ehrenfest dynamics (Figs.~\ref{fig:surf_dyn_1}e
and ~\ref{fig:surf_dyn_1}f) and may still be perceived as a fairly
good approximation.


{\it Breakdown of the friction model:} Considering the molecular model
in Fig.~\ref{fig:fric_fail}a with two possible initial energies, we
found qualitative break-down of the friction model. In a low-energy
regime (Fig.~\ref{fig:fric_fail}b) the friction trajectory experiences
a reflection at around $R = \SI{2}{\bohr}$. This reflection occurs
because the nuclear kinetic energy barely exceeds the ground-state
potential barrier and larger energy dissipation in the friction model
leads to the reflection. However, even with a large excess of energy,
a particle can still be trapped, as follows from
Fig.~\ref{fig:fric_fail}c. In this case we run into a resonance, when
the nuclear energy is quickly pumped into one of the oscillators
representing an electronic \gls{DOF}. Since the harmonic oscillators
can accept infinite amount of energy, trapping takes place. This
failure emphasizes the fact that the friction formalism even with the
full memory can give qualitatively incorrect results due to
intrinsically flawed representation of electronic \glspl{DOF} as
harmonic oscillators. On the other hand, dynamics in both
collective-mode models match very well that of the Ehrenfest method
for both setups.


In conclusion, we have derived a new \gls{MQC} dynamical model that is
fully \textit{ab initio}: the collective variables are constructed on
the fly from the nonadiabatic data available from electronic structure
calculations. The electronic dynamics in our approach is unitary and,
hence, properly accounts for electronic back-reaction and memory. Our
numerical simulations show that the collective variable approach
provides good and reliable description of nonadiabatic dynamics in a
variety of situations. The 2CM variant is accurate for ``an atom on an
insulator-like surface'' cases, whereas the 3CM model is applicable to
metallic surfaces as well. One of the collective variables coincides
with the ground-state amplitude, and can be used to draw a conclusion
whether the nonadiabatic effects are important or not. Comparison of
the collective-mode models with the full-memory \emph{ab initio}
electronic friction model shown the superiority of the former. The
employed friction model reflects a \emph{general} ability of any
friction model to reproduce nonadiabatic dynamics. Any additional
approximations to it may improve results only by fortuitous
cancellations of errors. The main drawback of the friction theory is
insufficient energy back-flow from the electronic subsystem. That can
cause nuclear reflection or trapping if certain and largely
unpredictable resonance conditions are met. Our observation supports
an idea that the similar failure reported in
Refs.~\citenum{Trail:2002/prl/166802}
and~\citenum{Trail:2003/jcp/4539} is caused by the friction approach
itself and is unrelated to the level of electronic structure theory.

{\it Acknowledgments: } A.F.I. acknowledges funding from a Sloan
Research Fellowship and the Natural Sciences and Engineering Research
Council of Canada (NSERC) through the Discovery Grants Program.

{\it Supporting Information: } The Supporting Information is available
free of charge on the ACS Publication website at DOI: xxx.

%

\end{document}